\documentclass[aps,prb,showpacs,twocolumn]{revtex4}
\usepackage{graphicx}
\usepackage{amsmath}
\usepackage{amsfonts}
\usepackage{color}

\begin{document}
\title{Andreev reflection in two-dimensional topological insulators with either conserved or broken time-reversal symmetry}

\author{Awadhesh Narayan and Stefano Sanvito}
\affiliation{School of Physics and CRANN, Trinity College, Dublin 2, Ireland}

\date{\today}

\begin{abstract}

We investigate Andreev reflection in two-dimensional heterojunctions formed by a superconductor in contact with a topological insulator 
ribbon either possessing or breaking time-reversal symmetry. Both classes of topological insulators exhibit perfect Andreev reflection, which is robust against disorder. This is assigned to topologically protected edge states. In the time-reversal symmetric case we show that doping 
one of the ribbon edges with magnetic impurities suppresses one Andreev channel, while no such suppression is seen in the broken symmetry 
situation. Based on this observation we suggest a tabletop transport experiment able to distinguish between the two types of topological insulators,
which does not involve the direct measurement of the material band structure.

\end{abstract}

\maketitle

\textit{Introduction}. Topological insulators (TIs) are a novel class of materials, which have been intensively investigated in the past few 
years.\cite{review-kane,review-zhang} They come in two varieties, $Z_2$ insulators, which preserve time-reversal symmetry, 
and Chern insulators, which do not. At present a variety of materials have been found to exhibit the time-reversal symmetric topological 
phase.\cite{zhangHgTe0,hasan-bisb,bi2se3} In contrast, although there are several theoretical proposals for constructing a Chern 
insulator,\cite{zhang-chern,niu-chern,zhang-HgMnTe} the experimental quest continues. 
An important question is how to design a measurement for distinguishing experimentally between these two classes of TIs. 
A direct probe of the band-structure is provided by spin-resolved angle-resolved photoemission spectroscopy (spin-ARPES),
which, however, is a complex experimental technique requiring a bright synchrotron light source. In this Rapid Communication we propose 
an alternative tabletop approach based on measuring the transport properties of an interface between a TI and a superconductor.

A metal/superconductor interface may reflect an incident electron from the metal as a positively charged hole with opposite spin, while 
a Cooper pair is formed in the superconductor. This electron-hole conversion is known as Andreev reflection \cite{andreev00} and has 
long served as a useful probe for spin-polarized currents.\cite{AndreevSpin} The Andreev reflection technique appears to 
be particularly suited to study edge state scattering in topological insulators and its heterostructures.\cite{cooperpair,sprox,nagaosa-majorana} 
Also intriguing is the possibility of interfacing $Z_2$ insulators with superconductors (SCs). This interface has been predicted to host 
Majorana fermions, with possible applications in topological quantum computing.\cite{kane-majorana} 

Recently there have been several interesting suggestions for turning the prototypical two-dimensional material graphene into a topological 
insulator. As far as the time-reversal symmetry broken case is concerned, Qiao \textit{et al.} have shown, by means of first principles 
calculations, that Fe atoms adsorbed on graphene open up a band-gap in the bulk and yield a non-zero Chern number for the valence 
band.\cite{niu-chern} In another ingenious proposal Weeks \textit{et al.} suggested to dope graphene with non-magnetic heavy adatoms 
in order to stabilize a time-reversal symmetric topological insulating phase.\cite{franz-qsh} In light of these promising developments, 
we have decided to study Andreev reflection in two-dimensional topological insulator-superconductor junctions.

\begin{figure}[h]
\begin{center}
  \includegraphics[scale=0.40]{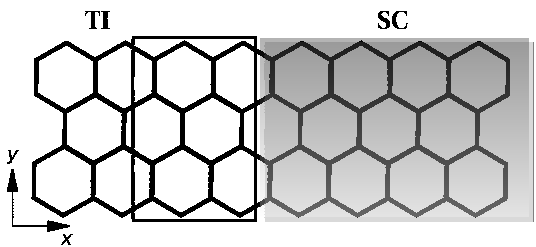}
  \caption{Setup for calculating the two-terminal transmission. Region SC is proximity coupled to a superconducting electrode while 
  region TI is the topological insulator described by the two chosen single-particle models. The rectangle marks the region at the TI/SC 
  interface where disorder is introduced.} \label{setup}
\end{center}
\end{figure}

\textit{Formulation}. We consider a two-dimensional topological insulator ribbon realized on a honeycomb lattice with zig-zag edge geometry, as shown 
in Fig.~\ref{setup}. The region to the right (SC region) is proximity coupled to a superconducting electrode, while the region on the left 
(TI region) is the topological insulator. The electron and hole spectra are described at the mean-field level by the Bogoliubov de Gennes 
equation,\cite{bdg}
\begin{equation}\label{bdg_hamiltonian} 
\begin{pmatrix}
H-E_{F}&\Delta\\
\Delta^{\ast}&E_{F}-{\cal T}H{\cal T}^{-1}
\end{pmatrix}
\begin{pmatrix}
u\\ v
\end{pmatrix}=
\varepsilon
\begin{pmatrix}
u\\ v
\end{pmatrix}\:,
\end{equation}
where $u$ and $v$ are the wave functions for electrons and holes, respectively. $H$ is the single-particle Hamiltonian for the topological 
insulators, $\cal T$ is the time-reversal operator, $\Delta$ is the pairing potential and $E_\mathrm{F}$ is the Fermi level. In the left region 
(TI) the pairing potential, is set to zero, i.e., there is no superconductivity. In the right region (SC) a finite constant pairing potential exists 
due to the proximity with a superconducting electrode. 

We use the Kane-Mele model \cite{kanemele} as single-particle Hamiltonian for the time-reversal symmetric ($Z_2$) topological insulator. 
This reads
\begin{equation}\label{kanemele}
H_\mathrm{KM}=t\sum_{<ij>}c_{i}^{\dagger }c_{j} + \lambda \sum_{i}\xi _{i}c_{i}^{\dagger }c_{i}+it_{2}\sum_{<<ij>>}\nu _{ij}c_{i}^{\dagger }\sigma^{z}c_{j}\:.
\end{equation}
Here the first term is just the nearest-neighbour hopping with strength $t$, where the spin indices of the creation, $c_{i}^{\dagger }$, and annihilation,
$c_{i}$, operators have been omitted. The second term represents a staggered sub-lattice potential, i.e., the $A$ type sub-lattice has an on-site 
energy $\lambda$ ($\xi=+1$), while the $B$ sub-lattice has on-site energy $-\lambda$ ($\xi=-1$). The last term describes second 
nearest-neighbour hopping with strength $t_{2}$ and it is purely imaginary ($t_2$ is real and $i=\sqrt{-1}$). Furthermore, $\nu_{ij}$ is equal to $+1$ 
for anti-clockwise hopping and to $-1$ for clockwise. Here $\sigma^z$ is the $z$-component Pauli matrix describing the electron's spin. The last term can be thought 
as a mirror-symmetric spin-orbit interaction, since it couples the orbital motion of the electrons to their spins.

For the time-reversal symmetry broken case we use a spinful version of the Haldane model \cite{haldane}, proposed by 
Chen \textit{et al.} \cite{spin-haldane} (from now on the spin-Haldane model, SH). The single-particle Hamiltonian reads
\begin{equation}\label{spinful-haldane}
 H_\mathrm{SH}=t\sum_{<ij>}c_{i}^{\dagger }c_{j} + \gamma\sum_{i}c_{i}^{\dagger }\sigma^{z}c_{i} + i\beta(\gamma)\sum_{<<ij>>}\nu _{ij}c_{i}^{\dagger }c_{j}\:,
\end{equation}
where the second term is the exchange field with strength $\gamma$ , i.e. it represents Zeeman coupling. In addition to spin, also the 
orbital angular momentum of the electron, $\nu_{ij}$, is coupled to the exchange field. Following Chen \textit{et al.}, \cite{spin-haldane} we approximate 
$\beta(\gamma)\approx \beta \mathrm{sgn}(\gamma)$, and choose $\beta$ to be negative. This parameter set describes a diamagnetic response 
to the magnetic field $\gamma$. Note that in this case the second nearest-neighbour hopping term has the same sign for both the spins, 
as opposed to that in $H_\mathrm{KM}$.

We use the ballistic Landauer-B\"{u}ttiker scheme \cite{buttiker} for calculating the transmission across the system. The self-energy 
matrix $\Sigma_\mathrm{L}$ ($\Sigma_\mathrm{R}$) for the left-hand side (right-hand side) contact is obtained by using the electrodes' surface 
Green's function, $g_s$. This is calculated iteratively from the following equation,\cite{datta}
\begin{equation}
 g_{s}=[(E+i0^{+})I-H_0-H_1 g_{s} H_1^{\dagger}]^{-1}\:,
\end{equation}
where $H_0$ is the Hamiltonian describing the electrode unit cell and $H_1$ is the coupling matrix between cells (note that in our tight-binding 
formulation the Hamiltonian of the ribbon has a tri-diagonal form). The self-energy matrices are given by $\Sigma_\mathrm{L, R}=
H_\mathrm{L, R} g_{s} H^{\dagger}_\mathrm{L, R}$, where $H_\mathrm{L}$ ($H_\mathrm{R}$) is the coupling matrix between the scattering 
region and the left-hand side (right-hand side) contact. Then, the retarded Green's function, $G^{r}$, for the scattering region of Hamiltonian
$H_\mathrm{SC}$ is obtained as $G^{r}=[(E+i0^{+})I-H_\mathrm{SC}-\Sigma_\mathrm{L}-\Sigma_\mathrm{R}]^{-1}$. 
The scattering region comprises the SC/TI interface and a portion of the electrodes. Finally the total transmission is simply 
$T(E)=\mathrm{Tr}(\Gamma_\mathrm{L}G^{r}\Gamma_\mathrm{R}G^{r\dagger})$, where $\Gamma_\mathrm{L, R}$ are the broadening 
matrices $\Gamma_\mathrm{L, R}=i[\Sigma_\mathrm{L, R}-\Sigma_\mathrm{L, R}^{\dagger}]$. Furthermore, the normal transmission 
coefficient from the $n$-th terminal to the $m$-th one is obtained as 
$T_{nm\sigma}(E)=\mathrm{Tr}(\Gamma_{n\sigma}G^{r}_{\sigma\sigma}\Gamma_{m\sigma}G^{r\dagger}_{\sigma\sigma})$, while the 
Andreev reflection coefficient is calculated as 
\begin{equation}\label{andreev_coeff}
R_{n\sigma,m\bar{\sigma}}^\mathrm{A}(E)=\mathrm{Tr}(\Gamma_{n\sigma}G^{r}_{\sigma\bar{\sigma}}\Gamma_{m\bar{\sigma}}G^{r\dagger}_{\bar{\sigma}\sigma}),
\end{equation}
where $\sigma=(\uparrow,\downarrow)$ and $\bar{\sigma}=(\downarrow,\uparrow)$ are the spin indices. Thus $R_{n\sigma,m\bar{\sigma}}^\mathrm{A}$ 
describes an incident electron from terminal $n$ being reflected as an opposite spin hole into terminal $m$.

\begin{figure}[h]
\begin{center}
  \includegraphics[scale=0.35]{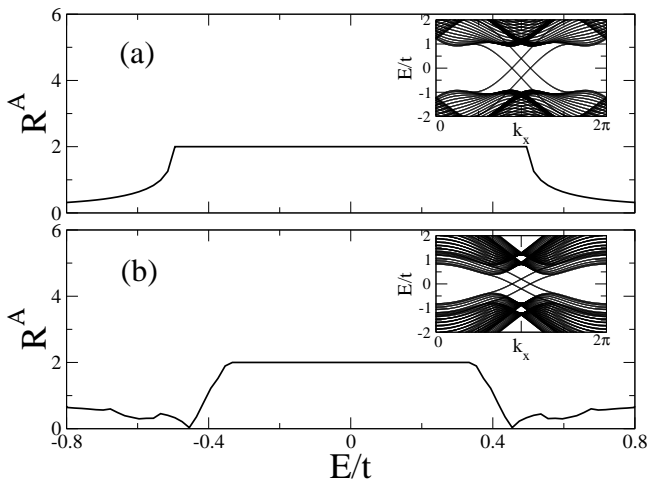}
  \caption{Andreev reflection coefficient for (a) $Z_{2}$ and (b) Chern insulators showing perfect Andreev reflection for electron 
  energies smaller than the superconducting gap. The insets show the bandstructure for the two models solved in a ribbon geometry. 
  Here we choose $t_{2}=0.33$, $\lambda/t_{2}=2.0$, $\gamma=0.20$, $\beta=-0.11$ and $\Delta=0.50$. The Fermi level $E_\mathrm{F}$ 
  is set at zero. }\label{andreev_pristine}
\end{center}
\end{figure}

\textit{Results}. We begin our analysis by calculating $R^\mathrm{A}$ as a function of energy, which is shown in Figs.~\ref{andreev_pristine}(a) 
and \ref{andreev_pristine}(b), respectively, for the Kane-Mele and spin-Haldane model. Henceforth, we set the nearest neighbor hopping $t=1$ and 
measure all the energies in units of $t$. Furthermore we fix the number of sites along the ribbon width to be $n_y=18$. The insets 
of Fig.~\ref{andreev_pristine} show the bandstructure for the two models calculated in this strip geometry. In the bulk gap there exist 
gapless edge modes, a single pair on each edge. For the $H_\mathrm{KM}$ Hamiltonian, these are opposite spin Kramer's pairs, 
while for $H_\mathrm{SH}$, there are two left movers (one for each spin) at one edge and two right movers at the other, exactly as the integer quantum Hall states. We find that for both these cases, the edge modes lead to a perfect Andreev reflection for electrons 
with energy smaller than the superconducting gap. In fact, in both cases the edge modes are perfectly Andreev reflected. 
Normal reflection, where an incident particle is reflected back without being converted into its antiparticle, is completely 
suppressed for the edge states as long as the Fermi energy lies in the bulk gap, as we have verified numerically. These findings are 
consistent with recent theoretical and experimental studies for time-reversal symmetric topological insulators.\cite{HgCdTe,InAsGaSb}
Note that by using a low-energy effective model for the edge states of a time-reversal symmetric TI, Adroguer {\it et al.} \cite{cooperpair} 
suggested Andreev reflection as a probe for helical edge states. Here we predict perfect Andreev reflection also for the time-reversal 
symmetry broken case.

Next, we study whether such perfect Andreev reflection is robust to perturbations of the electronic structure at the SC/TI interface. 
To this goal we consider the effect of onsite disorder, which is introduced by adding a term of the form 
$H_\mathrm{disorder}=\sum_{i}w_{i}c_{i}^{\dagger}c_{i}$ to both $H_\mathrm{KM}$ and $H_\mathrm{SH}$. Hence disorder enters 
in an exact and rather natural way in our numerical approach, at variance to low-energy edge models, where either a complex 
field theory construction or a perturbative treatment needs to be adopted. In particular here we choose the onsite energy, $w_i$, to be 
randomly distributed within the interval $[-W/2,W/2]$. Such disorder is introduced in a $n_x=15$ site-long region near the SC/TI interface. 

\begin{figure}[h]
\begin{center}
  \includegraphics[scale=0.3]{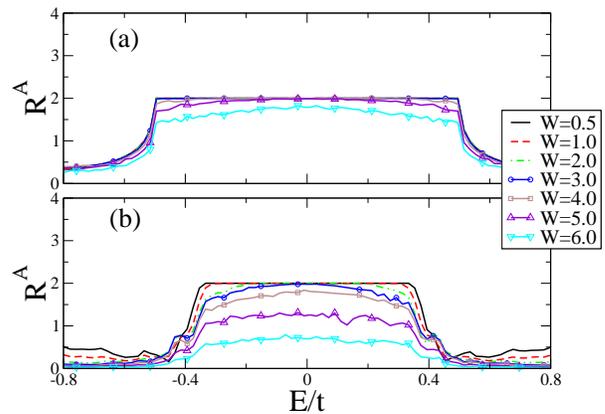}
  \caption{(Color online) Effect of onsite disorder on the Andreev reflection coefficient for (a) $Z_{2}$ and (b) Chern insulators. The 
  Andreev process is highly robust against onsite disorder and the crossover to dissipative transport occurs for $W\approx3.0t$ for the $Z_2$ insulator and $W\approx2.0t$ for the Chern insulator. Here again we set $t_{2}=0.33$, $\lambda/t_{2}=2.0$, $\gamma=0.20$, 
  $\beta=-0.11$ and $\Delta=0.50$, and the Fermi level $E_\mathrm{F}$ is taken at zero. The curves are averaged over 960 random 
  configurations. } \label{onsite_disorder}
\end{center}
\end{figure}

From Fig.~\ref{onsite_disorder} it can be clearly seen that the Andreev reflection process is very robust against disorder. Even for 
moderately large disorder ($W\approx2.0t$), $R^\mathrm{A}$ remains perfectly quantized. This is attributed to the presence of the 
topologically protected edge states, which are highly immune to impurities and disorder, and the situation is identical for both classes 
of topological insulators. For $W>2.0t$ fluctuations in $R^\mathrm{A}$ begin to develop in the energy range where only edge states 
exist. As a result the Fano factor becomes non-zero. This signals a transition from ballistic to diffusive transport where backscattering 
is allowed and the edge states are no longer topologically protected.\cite{buttiker-fano} Note that the actual value of the disorder strength 
critical for the destruction of the edge states depends on the robustness of the topological phase itself, i.e., on the model parameters used. 
However, as we will argue in what follows, the introduction of magnetic impurities breaks the topology of $Z_2$ insulators, even at weak 
disorder strengths, i.e., it is a general feature, which depends little on the model parameters.

This is demonstrated by introducing magnetic impurities at one of the TI ribbon edges. The exchange coupling between the electron 
spin and the impurities is incorporated into the model as  \cite{zhang-magneticimpurity} 
\begin{equation}
 H_\mathrm{ex}=\sum_{i, \alpha, \beta} c^{\dagger}_{i\alpha}\left[J_{z}\sigma_{z}^{\alpha\beta}S_{z}^i + 
 J_{\parallel}(\sigma_{x}^{\alpha\beta}S_{x}^i + \sigma_{y}^{\alpha\beta}S_{y}^i)\right]c_{i\beta}\:,
\end{equation}
where $S^i_n$ is $n$-th spin component of the magnetic impurity located at the edge site $i$, and $J_{z}$ and $J_{\parallel}$ are, respectively, the longitudinal and transverse exchange coupling. In the notation we have now explicitly re-introduced the spin index so that 
$c_{i\alpha}^\dagger$ ($c_{i\alpha}$) is the creation (annihilation) operator for an electron at site $i$ with spin $\alpha$. 
For simplicity here we have implicitly assumed that the magnetic impurities are not Kondo active. Their electronic structure is 
then treated at a simple classical level, i.e., they enter the model as classical spins.

When one includes only the $z$ component of the exchange coupling in the Kane-Mele model, there is a shift of the up and down spin 
edge bands, by an amount proportional to the coupling $J_z$ but no band gap opens in the edge state spectrum. For small values 
of $J_{z}$, before the bulk band gap closes, the system is in the time-reversal symmetry broken quantum spin Hall phase predicted by 
Yang {\it et al.}\cite{sheng-trsbrokenqsh} As a consequence, although we have locally broken time-reversal symmetry, perfectly quantized 
Andreev reflection still occurs. This is because in the energy range within the superconducting gap there is only the counter-propagating 
opposite spin channel available to normal reflection. 

\begin{figure}[h]
  \vspace{0.2in}
  \hspace{0.0in}
  \includegraphics[scale=0.35]{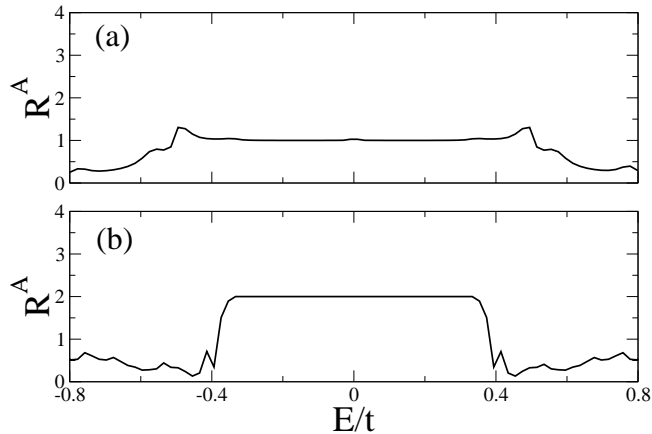}
  \caption{Andreev reflection coefficient in presence of magnetic impurities located at the edge of a TI ribbon: (a) $Z_{2}$ insulator 
  (b) Chern insulator. The suppression of one of the edge channels in the time-reversal symmetric case $R^\mathrm{A}$ produces a  
  drop in $R^\mathrm{A}$ from 2 to 1. Magnetic impurities have no effect on the Andreev reflection for a time-reversal symmetry broken 
  insulator. Here we have chosen $J_{z}=J_{\parallel}=0.50$ and $|S|=2$. The other parameters are the same as before. } \label{mag-imp}
\end{figure}

In contrast, if we also include the transverse component of the exchange, i.e., we take $J_{\parallel}\ne0$, then a gap is opened at the edge where 
the magnetic impurities have been located. The destruction of the helical edge states at one of the two edges leads to a suppression of 
this channel, which results in the Andreev reflection coefficient dropping from two to one, as shown in Fig.~\ref{mag-imp}(a). Such a 
reduction of $R^\mathrm{A}$ from 2 to 1 is almost perfect except for some bulk contributions at energies approximately equal to the 
superconducting gap. The situation for the spin-Haldane model is different and the magnetic impurities produce no effect, regardless of 
the magnitude of $J_{z}$ and  $J_{\parallel}$. This is expected, since the topological protection of the edge states for a Chern insulator 
continues to hold even in the absence of time-reversal symmetry. Consequently, no such suppression is observed and the Andreev 
reflection coefficient remains perfectly quantized to a value of two, as illustrated in Fig.~\ref{mag-imp}(b).

Thus, we have shown that Andreev reflection measurements can characterize a topological insulator and distinguish it from a topologically 
trivial material. Perfect Andreev reflection provides a signature for the existence of topological edge states, although it is not unique to them. 
One can in fact envisage other systems, which display a similar perfect electron hole conversion, for instance a pair of ballistic nanowires. What 
is unique, though, is the tremendous immunity to disorder, which both types of topological insulators display. Furthermore, we also showed 
that the $Z_2$ and Chern insulators respond differently to the presence of magnetic impurities. 

\begin{figure}[h]
  \vspace{0.2in}
  \hspace{0.0in}
  \includegraphics[scale=0.30]{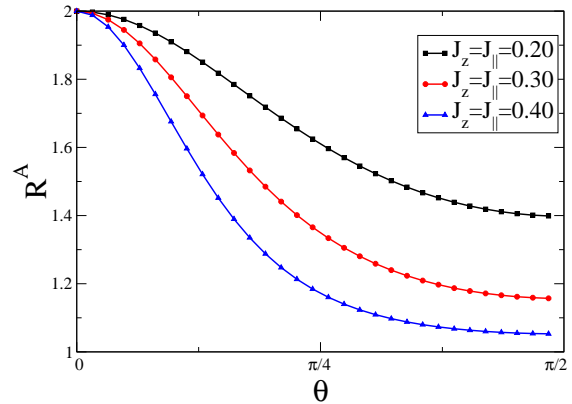}
  \vspace{-0.06in}
  \caption{ Andreev reflection coefficient in the presence of magnetic impurities for the $Z_{2}$ insulator as a function of the spin inclination angle 
  $\theta$ for various values of the exchange coupling. The relaxation of the spins leads to $R^\mathrm{A}$ reverting back towards unity.} \label{angular}
\end{figure}

Based on this observation we propose a transport experiment to distinguish between the two types of topological insulators. The experiment 
involves placing magnetic impurities along one of the edges of the two-dimensional sample, for instance, by using the tip of a 
scanning tunneling microscope. The impurities' spin will, in general, be aligned in arbitrary directions. The illumination with  
low-frequency polarized infrared light can however induce their alignment. This has been demonstrated, for instance, for Mn impurities in CdTe.\cite{awschalom-mn} The infrared pulse imparts a momentum to align the impurity spins, which subsequently relax back to their random 
orientations. The Andreev reflection coefficient $R^\mathrm{A}$ can then be measured as a function of time, and this can be related to the 
inclination angle $\theta$ of the impurity spin $S$. The dependence of $R^\mathrm{A}$ on $\theta$ is shown in Fig.~\ref{angular}. As the 
spin rotates towards the $z$ direction, $R^\mathrm{A}$ returns back to the perfectly quantized value of two, the same as that in the absence 
of impurities. For the Chern insulator the Andreev reflection process is unaffected by magnetic impurities and thus to the exposure with 
polarized light.

A second possible route to spin polarize the impurities consists in applying an intense static magnetic field $B$ perpendicular to the plane 
of the sample, and then to switch it off over a short time scale $t_{B}$. The impurity spins will then relax back to their random configuration 
with a typical spin-relaxation time $t_{S}$. During the time window comprised between $t_{B}$ and $t_{S}$, measurements of $R^\mathrm{A}$ 
should yield a behavior similar to that shown in Fig.~\ref{angular}. Note that these possible approaches were already outlined in Ref.[8] relative to the observation of the Chern insulating phase in Mn doped HgTe quantum wells. The same here are broadened in scope and now become a tool for assigning a given material to one of the two classes of TIs. Such a strategy mitigates the
need to perform a direct band-structure measurement, such as spin-ARPES, and therefore represents a powerful tabletop characterization 
method of the topological state of a material.

\textit{Summary}. In conclusion, we have investigated SC/TI  heterojunctions and shown that they display perfect Andreev reflection. The robustness of the
topologically protected edge states lends this effect a large immunity against disorder. We have then looked at magnetic impurities and 
shown that in the case of transverse exchange coupling the Andreev reflection coefficient of $Z_2$ topological insulators drops from two to
one. This observation allowed us to propose a transport experiment that is able to distinguish between the two types of topological insulators. 
This consists in following the time evolution of the Andreev reflection coefficient of a device dusted with magnetic impurities, which have been
previously polarized.

We would like to thank Ivan Rungger for helpful suggestions. This work is sponsored by the Irish Research Council for Science, Engineering 
and Technology (IRCSET) under the EMBARK initiative. Computational resources have been provided by the Trinity Centre for High 
Performance Computing (TCHPC).


\begin{thebibliography}{99}

\bibitem{review-kane}M.Z.~Hasan and C.L.~Kane, Rev. Mod. Phys. {\bf 82}, 3045 (2010).

\bibitem{review-zhang}X.-L.~Qi and S.-C.~Zhang, Rev. Mod. Phys. {\bf 83}, 1057 (2011).

\bibitem{zhangHgTe0}B.A.~Bernevig, T.L.~Hughes and S.C.~Zhang, Science {\bf 314}, 1757 (2006).

\bibitem{hasan-bisb}D.~Hsieh, D.~Qian, L.~Wray, Y.~Xia, Y.S.~Hor, R.J.~Cava and M.Z.~Hasan, Nature (London) {\bf 452}, 970 (2008).

\bibitem{bi2se3}Y.~Xia, D.~Qian, D.~Hsieh, L.~Wray, A.~Pal, H.~Lin, A.~Bansil, D.~Grauer, Y.S.~Hor, R.J.~Cava and M.Z.~Hasan, Nat. Phys. {\bf 5}, 398 (2009).

\bibitem{zhang-chern}R.~Yu, W.~Zhang, H.-J.~Zhang, S.C.~Zhang, X.~Dai and Z.~Fang, Science {\bf 329}, 61 (2010).

\bibitem{niu-chern}Z.~Qiao, S.A.~Yang, W.~Feng, W.-K.~Tse, J.~Ding, Y.~Yao, J.~Wang and Q.~Niu, Phys. Rev. B {\bf 82}, 161414(R) (2010).

\bibitem{zhang-HgMnTe}C.-X.~Liu, X.-L.~Qi, X.~Dai, Z.~Fang and S.C.~Zhang, Phys. Rev. Lett. {\bf 101}, 146802 (2008).

\bibitem{andreev00}A.F.~Andreev, Sov. Phys. JETP {\bf19}, 1228 (1964).

\bibitem{AndreevSpin} R.J.~Soulen {\it et al.}, Science {\bf 282}, 85 (1998).

\bibitem{cooperpair}P.~Adroguer, C.~Grenier, D.~Carpentier, J.~Cayssol, P.~Degiovanni and E.~Orignac, Phys. Rev. B {\bf 82}, 081303(R) (2010).

\bibitem{sprox}A.M.~Black-Schaffer, Phys. Rev. B {\bf 83}, 060504(R) (2011).

\bibitem{nagaosa-majorana}Y.~Tanaka, T.~Yokoyama and N.~Nagaosa, Phys. Rev. Lett. {\bf 103}, 107002 (2009).

\bibitem{kane-majorana}L.~Fu and C.L.~Kane, Phys. Rev. Lett. {\bf 100}, 096407 (2008).

\bibitem{franz-qsh}C.~Weeks, J.~Hu, J.~Alicea, M.~Franz and R.~Wu, Phys. Rev. X {\bf 1}, 021001 (2011).

\bibitem{bdg}M.~Tinkham, {\it Introduction to Superconductivity} (Dover, New York, 2004).	

\bibitem{kanemele}C.L.~Kane and E.J.~Mele, Phys. Rev. Lett. {\bf 95}, 146802 (2005).

\bibitem{haldane}F.D.M.~Haldane, Phys. Rev. Lett. {\bf 61}, 2015 (1988).

\bibitem{spin-haldane}T.-W.~Chen, Z.-R.~Xiao, D.-W.~Chiou and G.-Y.~Guo, Phys. Rev. B {\bf 84}, 165453 (2011).

\bibitem{buttiker}M.~B\"{u}ttiker, Phys. Rev. B. {\bf 38}, 9375 (1988).

\bibitem{datta}R.G.~Mojarad and S.~Datta, Phys. Rev. B {\bf 75}, 081301(R) (2007).

\bibitem{HgCdTe}Q.-F.~Sun, Y.-X.~Li, W.~Long and J.~Wang, Phys. Rev. B {\bf 83}, 115315 (2011).

\bibitem{InAsGaSb}I.~Knez, R.R.~Du and G.~Sullivan, arXiv:1106.5819v1 (unpublished).

\bibitem{buttiker-fano}M.~B\"{u}ttiker, Phys. Rev. Lett. {\bf 65}, 2901 (1990).

\bibitem{zhang-magneticimpurity}Q.~Liu, C.-X.~Liu, C.~Xu, X.-L.~Qi and S.-C.~Zhang, Phys. Rev. Lett. {\bf 102}, 156603 (2009).

\bibitem{sheng-trsbrokenqsh}Y.~Yang, Z.~Xu, L.~Sheng, B.~Wang, D.Y.~Xing and D.N.~Sheng, Phys. Rev. Lett. {\bf 107}, 066602 (2011).

\bibitem{awschalom-mn}D.D.~Awschalom, J.~Warnock and S. von Moln\'{a}r, Phys. Rev. Lett. {\bf 58}, 812 (1987).

\end{thebibliography}
\end{document}